# Competing Magnetic Interactions and Field-Induced Metamagnetic Transition in Highly Crystalline Phase-Tunable Iron Oxide Nanorods

Supun B. Attanayake[1], Amit Chanda[1], Thomas Hulse[2], Raja Das[3], Manh-Huong Phan[1], and Hariharan Srikanth[1,*]

[1] Department of Physics, University of South Florida, Tampa, Florida 33620, USA

[2] Department of Physics, University of Louisville, Louisville, KY 40208, USA

[3] SEAM Research Centre, South East Technological University, Waterford, Ireland

**Abstract**

The inherent existence of multi phases in iron oxide nanostructures highlights the significance of them being investigated deliberately to understand and possibly control the phases. Here, the effects of annealing at 250 $^0$C with a variable duration on the bulk magnetic and structural properties of high aspect ratio bi-phase iron oxide nanorods with ferrimagnetic $Fe_3O_4$ and antiferromagnetic $\alpha$-$Fe_2O_3$ is explored. Increasing annealing time under a free flow of oxygen enhanced the $\alpha$-$Fe_2O_3$ volume fraction, and improved the crystallinity of the $Fe_3O_4$ phase, identified in changes in the magnetization as a function of annealing time. A critical annealing time of approximately 3 hours maximized the presence of both phases, as observed via an enhancement in the magnetization and an interfacial pinning effect. This is attributed to disordered spins separating the magnetically distinct phases which tend to align with the application of a magnetic field at high temperatures. The increased antiferromagnetic phase can be distinguished due to the field-induced metamagnetic transitions observed in structures annealed for more than 3 hours and was especially prominent in the 9-hour annealed sample. Our controlled study in determining the changes in volume fractions with annealing time will enable precise control over phase tunability in iron oxide nanorods, allowing custom-made phase volume fractions in different applications ranging from spintronics to biomedical applications.

*Corresponding author: sharihar@usf.edu

# 1. Introduction

Since the groundbreaking revelation of the concept in the 1960s by Richard Feynman, nanoparticles have made significant strides in many fields, especially magnetic nanoparticles have attracted significant attention for their potential applications in drug delivery, cancer treatment, and hyperthermia.[1–7] The scope of these fields has recently expanded into spintronic devices as well, which cover a wide variety of applications such as storing, processing, and transmitting data as well as in the energy sector showing potential applications as effective catalysts in fuel cells[8–14] Iron oxide nanoparticles which are discussed here are well-known in all of the above-mentioned fields as they are non-toxic, stable, able to functionalize and possess a high surface-volume ratio.[15–18] Additionally the presence of multiple phases of iron oxide such as α-$Fe_2O_3$, γ-$Fe_2O_3$, $Fe_3O_4$, etc. possess distinct magnetic ordering. These phases enable iron oxide nanoparticles and thin films to be customized to variable applications.[19–25]

High aspect ratio nanorods synthesized by Das *et al.*[26] illustrated exotic magnetic and inductive heating properties mainly due to the high surface-volume ratio and anisotropy. The epitaxial growth of these highly oriented $Fe_3O_4$ nanorods on $SrTiO_3$ substrates has created a novel heterostructure with enhanced perpendicular magnetic anisotropy, which is desirable for magnetic data storage and spintronic devices.[12] Functionalizing the structure not only enhances its basic capabilities but opens a wide range of additional applications. For example, usage as a contrast agent in magnetic resonance imaging, as a gas sensor, or as an anode in a battery cell.[27–30] To further extend the functionalization of iron oxide nanorods, phase, and structural tunability are necessary. A rigorous examination of iron oxide nanorods via magnetometry enabled Attanayake *et al.*[19] to determine the presence of biphase which is undetectable when using orthodox characterization methods such as X-Ray Diffractometry (XRD). These results proposed the opportunity to study the phase tunability and coexistence of antiferromagnetic (AFM) α-$Fe_2O_3$ and ferrimagnetic $Fe_3O_4$. α-$Fe_2O_3$ is thermodynamically more stable and its changes with varying annealing temperatures in the presence of free-flow of oxygen were initially checked to determine an ideal formation temperature, which preserves the magnetic qualities of $Fe_3O_4$ but leverages the characteristics of α-$Fe_2O_3$.[30–32] This temperature was later determined to be 250 $^0$C and has been utilized in all of the annealing runs in the current experimental setup which was annealed at varying durations in the presence of free-flow of oxygen to observe the phase changes as a function of annealing time.

In this context, we report on the results of a thorough study of the magnetic and structural properties of high aspect ratio iron oxide nanorods with ferrimagnetic $Fe_3O_4$ and antiferromagnetic α-$Fe_2O_3$ phase tunability. A comprehensive understanding of the phase changes in these iron oxide nanorods will allow for the precise tailoring of these structures, enhancing the phase-specific magnetic and structural properties and giving greater control over the phase-tunability of biphase iron oxide nanorods paving the way for a wider range of potential applications.

## 2. Experimental Details

The iron oxide nanorods (NR) were synthesized by closely following the method proposed by Sun *et al.*[33] The initial solution containing 0.4 g of Hexadecylamine(HDA), 4 ml of Oleic acid(OA), and 16 ml of 1-Octanol were continuously stirred for 30 minutes at 55 $^0$C temperature. The ratio of HDA to OA which determines the aspect ratio of the NR was fine-tuned by varying the amount of HDA, based on the work of Raja *et al.*[26] The resulting clear solution was then left to reach room temperature, into which 4 ml of Iron(0) Pentacarbonyl was added and stirred for another hour. The solution in a Teflon-lined container was then encased in a steel metal jacket and placed inside a 200 $^0$C pre-heated oven for 6 hours to undergo autonomous pressure resulting in elongated nanostructures. After the solution reached room temperature, the dark-colored nanostructures were retained in the solution while the yellowish supernatant was poured out. The remainder of the solution was then cleaned using ethanol and a small amount of hexane by sonicating and centrifuging and repeated at least two times or until the solute is well separated. The well-separated product is then left to dry for a minimum of three days to obtain a fine powder. The powder is then placed in a ceramic combustion boat and inserted into a tube furnace heated to 250 $^0$C with a continuous flow of oxygen. The samples were kept at this temperature for durations of 1, 3, 5, 7, and 9 hours, and finally, all the samples were examined for their structural and compositional consistency by respectively using the FEI Morgagni 268 Transmission Electron Microscope (TEM) and Bruker AXS D8 X-Ray Diffractometer (XRD). All the magnetic measurements were completed by using the Vibrating Sample Magnetometer (VSM) option in a DynaCool Physical Property Measurement System (PPMS) manufactured by Quantum Design USA.

## 3. Results and Discussion

Fig. 1 shows how the iron oxide nanorods changed their structural composition with annealing and additionally the inset in Fig. 1(a) shows the TEM image of a high-quality as-prepared sample (AP) with low agglomeration, uniform shape with a size distribution of 35 nm and an aspect ratio of ~6.

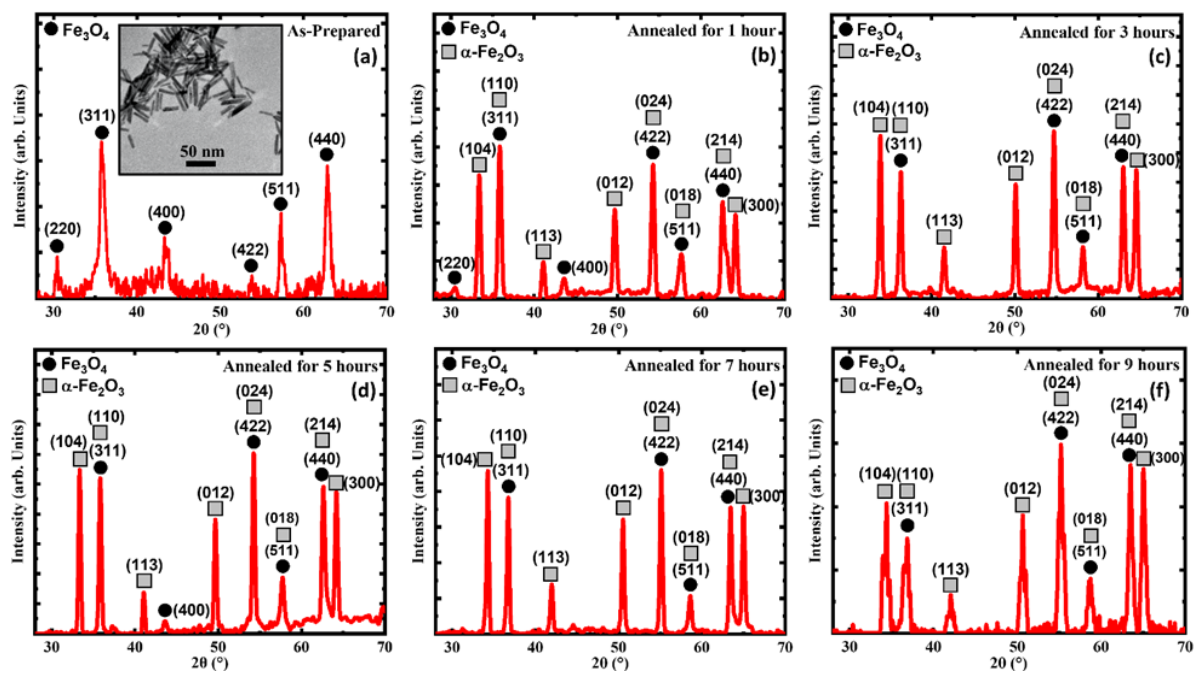

**Figure 1:** XRD patterns of the (a) as-prepared iron oxide nanorods (inset shows a TEM image of the as-prepared iron oxide nanorods), (b) 1 hour annealed, (c) 3 hours annealed, (d) 5 hours annealed, (e) 7 hours annealed, and (f) 9 hours annealed.

A closer look at the TEM image shows that the structures are less sharp which hints that the AP has not reached its full crystallinity, which is further confirmed via the XRD data. The AP which was annealed for durations of 1, 3, 5, 7, and 9 hours at 250 $^0$C show the presence of the additional iron oxide phase of the α-$Fe_2O_3$ phase, which was not visible due to the capping provided by the Oleic Acid. The sharpness of the XRD peaks in the annealed samples (AN) is comparatively higher than the AP, which hints at the increased crystallinity with annealing.[34] Furthermore, the AN showed the consistent appearance of biphase in the AP with annealing.

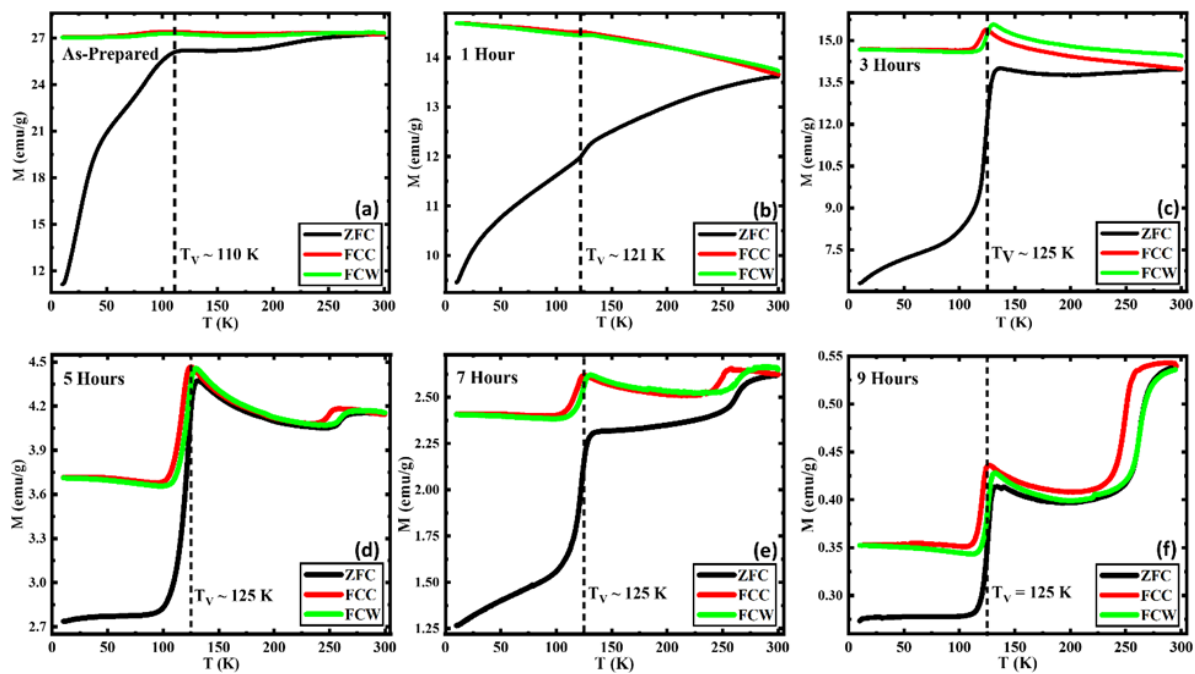

**Figure 2:** ZFC and FC M(T) curves in an applied field of 0.05 T for (a) as-prepared, (b) annealed for 1 hour, (c) annealed for 3 hours, (d) annealed for 5 hours, (e) annealed for 7 hours, and (f) annealed for 9 hours iron oxide nanorods.

The temperature-dependent magnetization M(T) curves were measured following the zero-field cooled (ZFC), field-cooled (FC), and field-cooled-warming (FCW) protocols in presence of a static magnetic field of 0.05 T. Fig. 2(a) which depicts a significant slope change below ~50 K is understood to be a common phenomenon in many of the systems, specifically in ferrite nanoparticle systems.[35–37] The prominent magnetic shoulder found below ~50 K was confirmed via AC susceptibility measurements by Bhowmik *et al.* which showed that such features are due to strong magnetic spin interactions and defects in the lattice structure further enhancing the randomness in the surface spins.[21,38–41] The bifurcation between the ZFC and FC curves known as the blocking temperature is observed only in the AP below 300 K since the elevated AFM phase has increased the blocking temperature past room temperature.[42,43] The Verwey transition (VT) at which the high-temperature cubic spinel transforms into a low-temperature monoclinic structure, is a significant transition of $Fe_3O_4$, observed in all the samples between 110 K and 125 K, which can generally lie in the range of 80-125 K.[44,45] The change in the crystallographic structure is accompanied by electrical resistivity, heat capacity, magnetic susceptibility, remanence, and coercivity ($H_C$) changes around the Verwey transition temperature ($T_V$).[44–47] The $T_V$ tends to increase with the crystallinity, plateauing at approximately 125 K. This first-order metal-insulator transition reached its peak after annealing it for 3 hours at 250 $^0$C. This implies that the samples require at least 3 hours of exposure to a continuous flow of oxygen at 250 $^0$C to reach their full crystallinity[44,48]. Furthermore, the sharp, first-order transition in samples 3, 5, 7, and 9 indicates that the $Fe_3O_4$ phase is with sufficient purity and/or stoichiometry. This is indicative of the increasing crystallinity of the $Fe_3O_4$ phase with the annealing time, and vice-versa the multistage transitions below 125 K which are less sharp compared to the prior. This can be understood to be due to less crystallinity/stoichiometry in the $Fe_3O_4$ phase in the AP and 1-hour AN.[48] Additionally, the suppression of the VT can occur due to slight oxidation of the $Fe_3O_4$ phase into a $\gamma$-$Fe_2O_3$-like phase which has been observed in capped nanoparticle systems.[49] The Morin transition (MT), which is a hallmark transition associated with the $\alpha$-$Fe_2O_3$ (hematite) phase, can be observed in almost all the samples other than the 1 and 3-hour AN. The MT, commonly known as the temperature-driven spin flop transition and is ideally associated with a first-order magnetic phase transition in a $\alpha$-$Fe_2O_3$ where it transforms from a weak ferromagnetic phase with spins aligned perpendicular to the c-axis above the MT to a fully AFM phase with spins aligned parallel to the c-axis below MT along with the change of sign of the magnetic anisotropy constant.[50–55] The absence of the MT in the 1 and 3-hour AN can be due to several reasons. The MT which occurs due to a subtle change of the orientation of spins in $\alpha$-$Fe_2O_3$ can get affected due to impurities, defects, spin frustrations, etc. in the structure and can lead to the diminishing or disappearance of the MT.[56] The concurrent occurrence of the Morin and Verwey transitions, respectively in samples other than 1 and 3-hour annealed samples indicates the coexistence of the α-$Fe_2O_3$ and $Fe_3O_4$ phases in these samples, which was also confirmed by the XRD analysis. The α-$Fe_2O_3$ volume fraction has drastically increased in the samples that are annealed for a longer duration, as identified by the enhanced sharpness of the respective MTs. Prominent thermomagnetic hysteresis has been observed between the field-cooled cooling (FCC) and field-cooled warming (FCW) M(T) for the 5,7, and 9 hours AN and it becomes stronger in the 9 hours AN compared to the other two, which also complements our hypothesis about enhanced volume fraction of the α-$Fe_2O_3$ phase with increasing annealing duration. The significant separation between the FCC and FCW in Fig. 2(c) indicates that the sample was able to achieve a higher magnetization at the end of FCW compared to the magnetization achieved at the start of FCC. With the spins aligned at the start of the FCC protocol, the magnetization further increases as the temperature drops since the magnetic moments align with the external field. While heating up when following the FCW protocol, the magnetic moments which did not possess enough energy to align, further aligned with the applied field leading to a separation between the FCC and FCW magnetization curves.

In Fig. 3, we show the magnetization vs. the applied field M(H) at room temperature for all the samples. It is noteworthy that the magnetization at the highest magnetic field 1 T ($M_S$), has dropped with annealing in all the samples except for the 3 hours AN, which shows an enhancement in its magnetization. This enhancement is understood to be due to an optimization of phase coexistence and crystallinity. The AP showed the highest $M_S$ along with the lowest coercivity $H_C$ and room temperature superparamagnetic behavior indicating the sample acts as a single domain at room temperature.[57] Furthermore, the AP showed a decent approach to saturation while the others did not show full saturation even at higher fields indicating the increased volume fraction of the AFM α-$Fe_2O_3$ phase since ideally to achieve the saturation magnetization of a FiM/AFM system the applied field needs to surpass the spin-flop transition which is in 10-100 T in magnitude.[58–60]

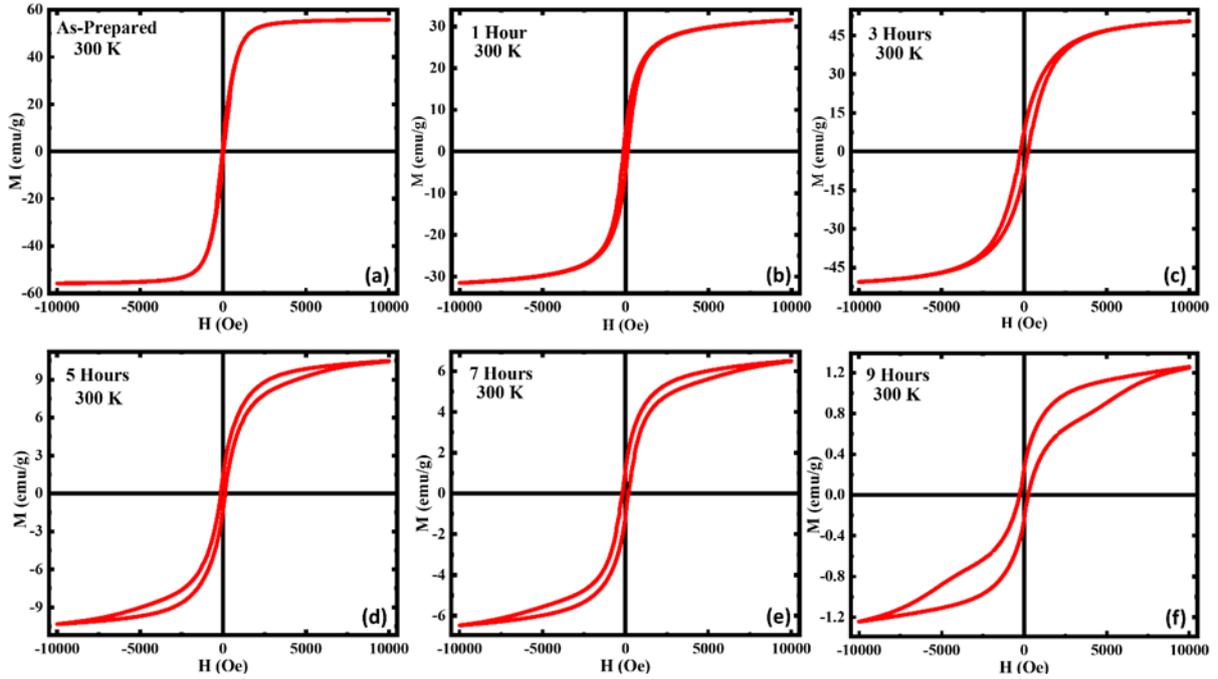

**Figure 3:** Magnetic hysteresis loops M(H) of (a) as-prepared, (b) annealed for 1 hour, (c) annealed for 3 hours, (d) annealed for 5 hours, (e) annealed for 7 hours, and (f) annealed for 9 hours iron oxide nanorods at 300 K.

This increase in the volume fraction of the AFM phase can be further observed with the decrease in the $M_S$ compared to the AP. The $H_C$ has increased with annealing and 3-hours annealed shows the highest. The increase in $H_C$ can be attributed to the strong competition between the AFM α-$Fe_2O_3$ phase, and the magnetically softer FiM $Fe_3O_4$ phase in the annealed samples. Furthermore, the decrease in magnetization values can also be associated with the enhanced volume fraction of the AFM α-$Fe_2O_3$ phase. Most interestingly, for the 5,7, and 9 h AN, the increasing branch of the isothermal M(H) loops exhibit a slope change around 5 kOe and this behavior is more robust in the 9 h AN. The magnetization of the increasing branch of the 9 h AN shows a sudden but smooth jump above the slope change which indicates the occurrence of field-induced metamagnetic transition in this sample.[61] We believe that the appearance of this field-induced metamagnetic transition is associated with the presence of the AFM α-$Fe_2O_3$ phase. Field-induced metamagnetic transition usually occurs when a

ferromagnetic(FM)/FiM phase exists with an AFM phase and the occurrence of this phenomenon in the annealed samples (for durations above 5 h) can be due to the superficial spin disorder with the mixed magnetic phases and/or core-shell-like structure.[62] With the application of a magnetic field, the AFM phase reorients with FiM order giving rise to the metamagnetic transition.[63]

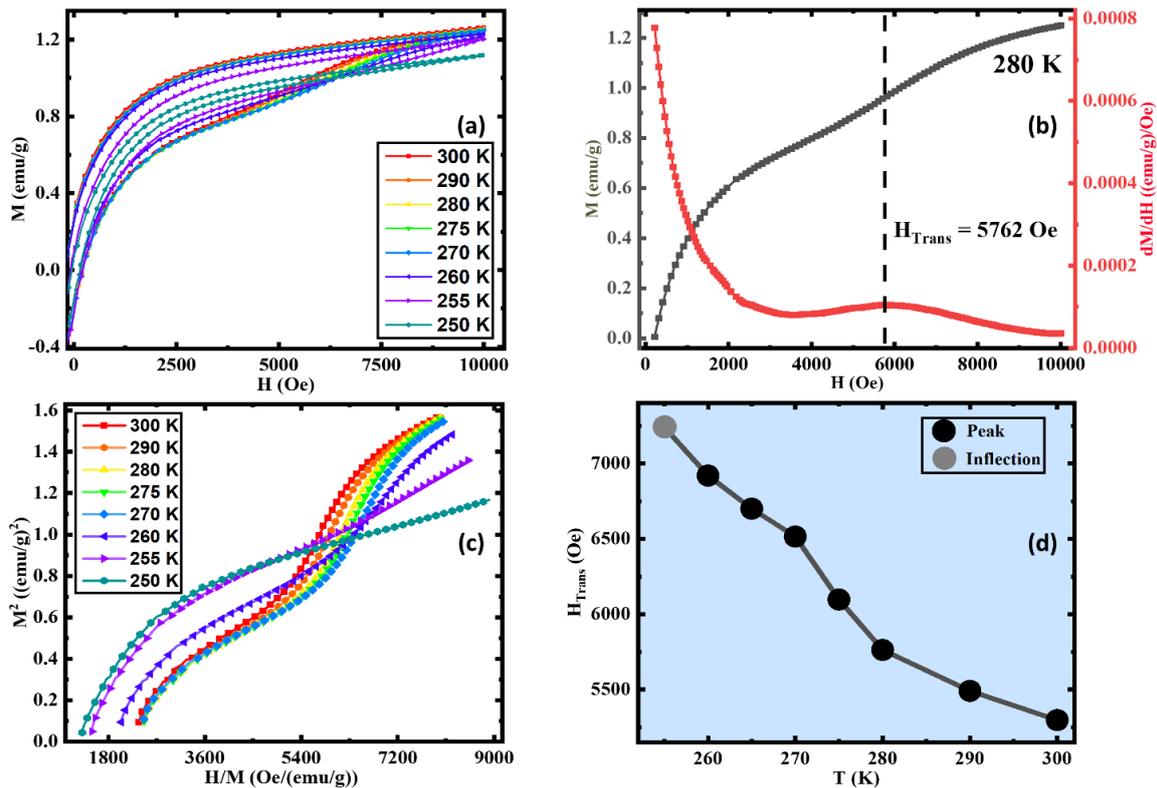

**Figure 4:** (a) Magnetic hysteresis loops M(H) between 250 K and 300 K, (b) magnetization curve and the magnetization with respect to field curve at 280 K, (c) Arrott curves between 250 K and 300 K, and (d) the transition field ($H_{Trans}$) vs temperature of the 9 hours annealed sample

For a clearer understanding, we performed M(H) measurements on the 9 hours AN sample at selected temperatures above and below the Morin transition. As shown in Fig. 4(a), the slope change in M(H) associated with the metamagnetic transition becomes stronger with decreasing temperatures up to ~275 K, below which it slowly weakens and disappears at T = 250 K. Usually, the metamagnetic transitions are the first order transition. To understand the order of the field-induced metamagnetic transitions in our AN sample, we show Arrott plots for 9 h annealed samples at selected temperatures in the range 250 K ≤ T ≤ 300 K in Fig. 4(c). In the case of a first-order metamagnetic transition, the fourth-order coefficient of the Landau free-energy expansion becomes negative giving rise to the appearance of an inflection point/negative slope accompanied by an S-shaped curve of the Arrott plot.[64,65] It is evident that the Arrott plots for our 9 h AN exhibit S-shaped curves along with inflection point and the negative slope in the field range 5400 Oe ≤ H ≤ 7200 Oe at temperatures in the range 260 K ≤ T ≤ 300 K which further confirms the occurrence of field induced first-order metamagnetic transition in this sample.[66] The S-shape behavior of the Arrott plots disappears for T ≤ 255 K. We estimated the transition field associated with the metamagnetic transition ($H_{Trans}$) from the first derivative of the increasing branch of the isothermal magnetization curve, as indicated in Fig. 4(b). Clearly, $H_{Trans}$ shifts to higher fields with decreasing temperature and eventually disappears below 255 K, as shown in Fig. 4(d).

In Fig. 5(b), we show the isothermal M(H) loops for all the samples at T = 10 K measured under zero-field cooled (ZFC) and field cooled (FC) protocols with the cooling field of $\mu_0 H$ = 1 T. It is evident that the magnetization value at the highest applied magnetic field (1T) remains the same for ZFC and FC M(H) loops for all the samples. Interestingly, the FC M(H) shows an enhancement magnetization in the low field region (below ~5 kOe) for all the samples in comparison to the ZFC M(H) except for the 1 h AN sample for which the aforementioned low field enhancement is comparatively smaller than the other samples. Furthermore, the enhancement in the low field magnetization in the FC M(H) loops becomes more robust as the annealing duration increases. However, we have not observed any horizontal/vertical shifts and/or increase in $H_C$ in the FC M(H) loops relative to the ZFC M(H) loops for any of these samples indicating the absence of exchange bias (EB) effect in these samples.

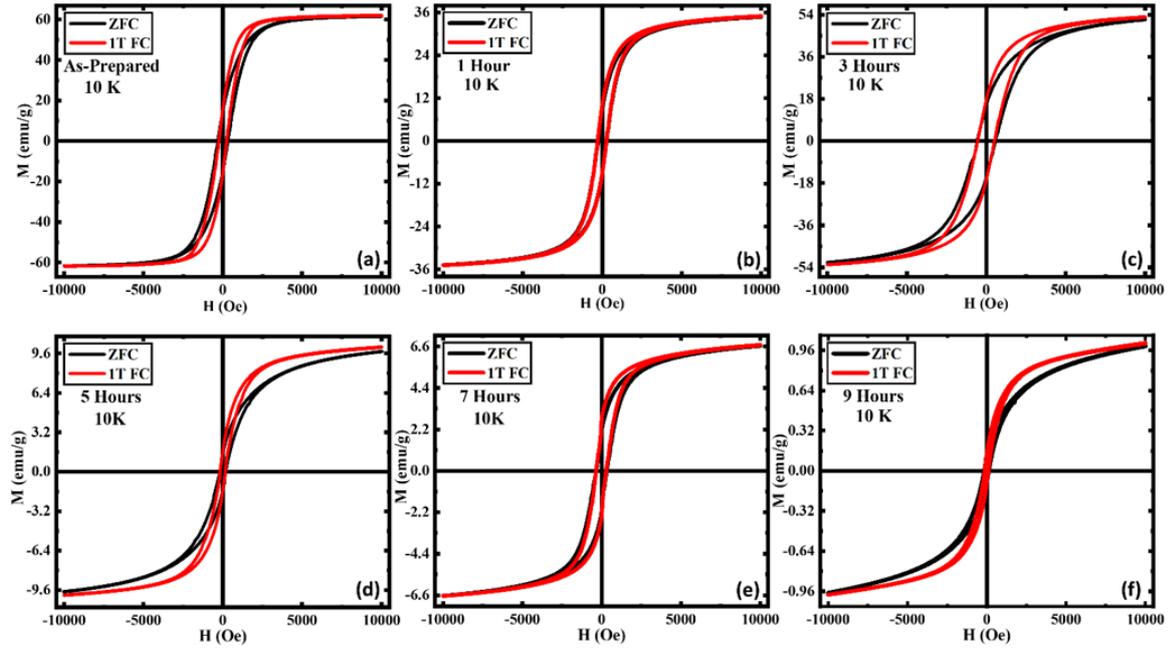

**Figure 5:** Magnetic hysteresis loops M(H) recorded with ZFC and FC protocols in an applied field of 1T for (a) as-prepared, (b) annealed for 1 hour, (c) annealed for 3 hours, (d) annealed for 5 hours, (e) annealed for 7 hours, and (f) annealed for 9 hours iron oxide nanorods at 10 K.

Such low field enhancement in magnetization and the absence of exchange bias effect in these samples can be interpreted in terms of the presence of weakly ordered AFM α-$Fe_2O_3$ phase.[59] In our bi-phase iron oxide system, the ferrimagnetically ordered $Fe_3O_4$ phase is the dominating phase the volume fraction of which is getting replaced by the AFM α-$Fe_2O_3$ phase with increasing annealing duration. However, the sharpness of the Verwey transition strongly indicates the existence of the highly crystalline $Fe_3O_4$ phase along with the AFM α-$Fe_2O_3$ phase even in the 9 h annealed sample. At low temperatures, the total magnetic moment associated with the $Fe_3O_4$ phase contributes significantly towards total magnetization, even if the volume fraction of the aforementioned phase is small compared to the AFM counterpart of the system. Therefore, the AFM interaction among the spins associated with the AFM α-$Fe_2O_3$ phase is likely to be destabilized by the exchange field of the FiM $Fe_3O_4$ phase and hence, weakens the AFM ordering of the α-$Fe_2O_3$ phase.[59] Usually, a large magnetic field (much above the spin flop transition field of the AFM, which is typically in the range of 10-100 T) is needed in order to fully saturate a classical AFM material. However, it has been recently shown that the spin-flop transition field for the α-Fe$_2$O$_3$ thin films is much lower (~ 3 kOe) than conventional classical AFMs,[60] indicating that the macro-spins associated with the AFM α-Fe$_2$O$_3$ can be easily saturated by applying 1 T magnetic field. In this scenario, the spins associated with the weakly ordered AFM α-Fe$_2$O$_3$ phase remain unaligned in the low field region when the sample is cooled in a zero applied magnetic field and leads to a lower value of magnetization.[59] However, the spins associated with the weakly ordered AFM α-Fe$_2$O$_3$ phase get aligned along the field direction when the sample is cooled in a higher magnetic field (1 T) and gives rise to the enhanced magnetization in the low field region.

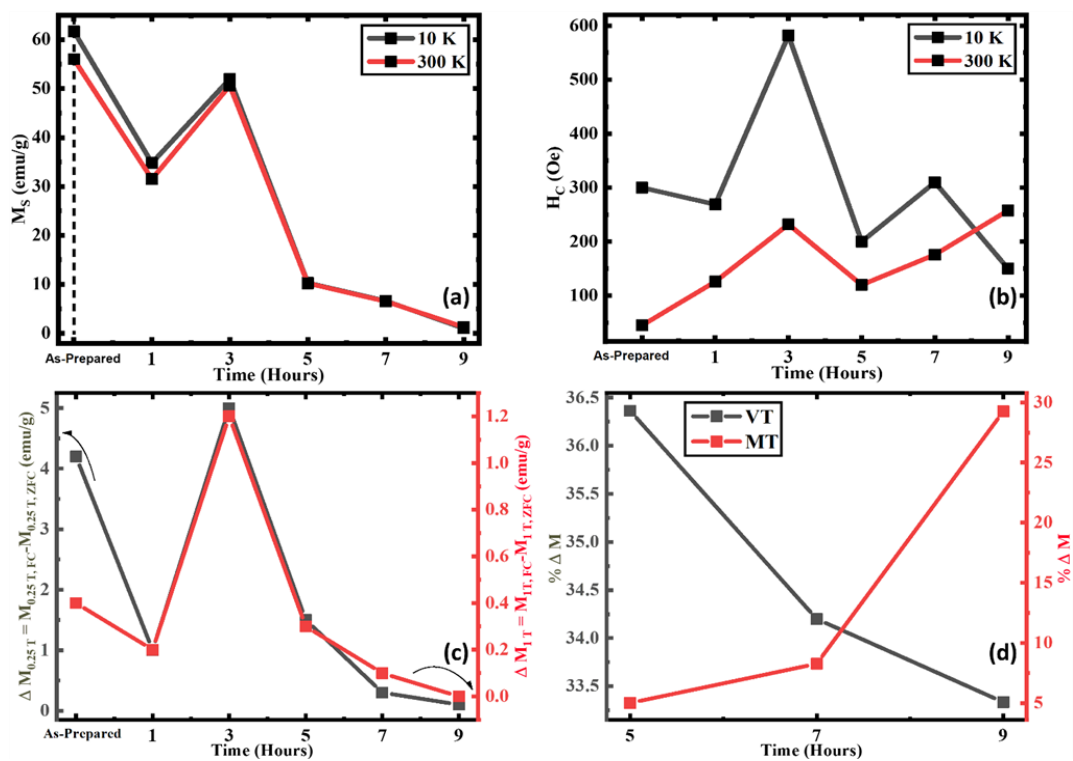

**Figure 6:** (a) Maximum magnetization at 1 T, (b) Coercivity at 10 K and 300 K, (c) Difference between the FC and ZFC magnetization at 10 K on 0.25 T and 1 T applied field of iron oxide nanorods, and percentage change in magnetization around VT and MT of 5, 7, and 9 hour annealed iron oxide nanorod samples.

As shown in Fig. 6(a), the $M_S$ value (magnetization at the highest magnetic field 1 T) at 10 K and at 300 K show variation between their respective low and high-temperature values which can be observed in the AP, 1-hour, and 3-hour AN. The $M_S$ tends to be higher at a low temperature in comparison to a high temperature as the magnetic moments can be easily aligned with fewer thermal fluctuations.[67–69] Clearly, the $M_S$ value for the 9-hour AN sample is minimum amongst all the other samples at both 10 and 300 K. Moreover, the difference in the $M_S$ values between 10 K and 300 K significantly decreases with an increase in the annealing duration. All these observations strongly indicate enhancement in the volume fraction of the AFM α-$Fe_2O_3$ phase with increasing durations. This can be compared with Fig. 6(b) where the $H_C$ at low and high temperatures followed the same trend, until the 9-hour AN where the weakly FM α-$Fe_2O_3$ phase above the MT results in a higher $H_C$ at the high temperature. The enhancement of the $H_C$ until the 3-hour AN in both temperatures is understood to be due to the interfacial interaction between the AFM and FiM phases with the optimization of the two phases, as the spin distribution and the arrangement change.[70] Specifically, the significant enhancement of the crystallinity in the FiM $Fe_3O_4$ phase of the 3-hour AN results in a better ordering of spins which has a higher potential in influencing the spins at the interfaces as they prevent the spins to get freely ordered with the applied field leading to the enhanced $H_C$.[71] When the AFM phase further increases the $H_C$ decreases as the interfacial interactions get directly affected by the effective area and microstructural change between the two phases as the local spin frustrations result.[70,72] Fig. 6(c), calculated using:

$$\Delta M_X = M_{X, FC} - M_{X, ZFC} \qquad (1)$$

where $\Delta M_X$ denotes the difference in magnetization at the "X" applied field. Here the relation highlights the difference between the FC and ZFC magnetization at 0.25 T and 1 T applied field points which clearly shows that at around 0.25 T field, the difference between the FC and ZFC magnetizations is higher, indicating the pinning effect.[73] Though the trend is similar at both the said applied fields of 0.25 T and 1 T, the AP sample is an exception as it shows a higher change in $M_S$ values at 0.25 T which can be understood to be due to the significantly higher FiM $Fe_3O_4$ with the low AFM $\alpha$-$Fe_2O_3$ phase. This is the reason we observe a larger difference between the magnetization in the FC and ZFC protocols in the low-field field (0.25 T) as the $\alpha$-$Fe_2O_3$ phase aligns.[59] The percent change in magnetization (% $\Delta M$) at VT and MT has been calculated by using the:

$$\% \ \Delta M = (M_{max} - M_{min})/M_{max} \qquad (2)$$

and considering the significant step change of the ZFC curve at ~VT and ~MT selected by the higher and lower inflection points in Fig. 2(d), 2(e), and 2(f). Fig. 6(d) shows a decrease in the percentage change in magnetization at the VT while simultaneously it increases at the MT as the annealing duration increase. This implies that the volume fraction of the AFM $\alpha$-$Fe_2O_3$ phase increases with the annealing duration. The percentage change in the magnetization at ~VT though significant is quite low as the FiM $Fe_3O_4$ phase has attained its peak crystallization at 3 hours of annealing and the decrease is solely corresponding to the decrease of the FiM $Fe_3O_4$ volume fraction.

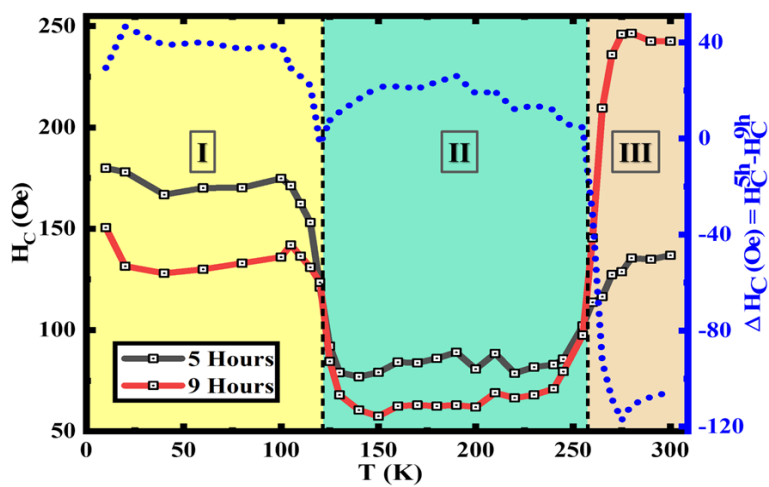

**Figure 7:** Temperature dependence of coercivity and coercivity difference between the 5-hour annealed and 9-hour annealed iron oxide nanorods.

Here, the temperature dependence of $H_C$ of the 5-hour annealed and the 9-hour annealed samples are extracted from M(H) loops at varying temperatures, these clearly show the VT and the MT at ~120 K and ~260 K respectively. As the temperature is increased from 10 K the $H_C$ initially drops from a high value at around ~50 K as this has been observed to be the spin-freezing temperature in many of the nanoparticle systems.[19,20,35,74] The increased $H_C$ below the VT in *region I*, is understood to be due to the large magneto crystalline anisotropy change which occurs at the VT. Here the anisotropy decreases with the temperature increment across VT, as the crystalline structure changes from monoclinic to cubic spinel, which can be observed in both samples.[75–77] The higher change of $H_C$ was observed in the 5-hour annealed sample as the volume fraction of $Fe_3O_4$ is higher, it leads to a larger change in magneto-crystalline anisotropy across the *region I* and *II*. The biphase nanostructures between *regions I* to *II* show decreased $H_C$ as the anisotropy drops above VT in $Fe_3O_4$. The increase in $H_C$ at the MT when approaching *region III* indicates the anisotropy change associated with the MT. Here the 9-hour annealed sample shows a much higher $H_C$ change across MT as it possesses structures with a higher volume fraction of $\alpha$-$Fe_2O_3$ which transitioned from AFM to weakly FM when moving across MT. This phenomenon is related to the anisotropy change which is resulted from the change in magnetic moments aligning along the c-axis to perpendicular to it.[19,51,78] The difference between the $H_C$ values depicted by the right-hand axis in Fig. 7 clearly shows how the $H_C$ of the 9h-annealed sample dominates after passing the MT in *region III*.

## 4. Conclusions

A systematic study of the annealing-induced phase evolution of iron oxide nanorods was conducted at 250 $^0$C for varying durations of time, enabling a closer look at how the crystallinity and the volume fractions of $Fe_3O_4$ and $\alpha$-$Fe_2O_3$ can be tuned between major and minor phases. The pinning effect was observed in all the samples except the 1-hour annealed sample which is understood to be due to the γ-$Fe_2O_3$-like phase inhibiting the pinning effect under the low-temperature field-cooled protocol. The disappearance of the Morin transition despite the XRD results was understood to be due to strong magnetic interactions along with impurities/defects. The annealing done at 250 $^0$C for 3 hours showed enhanced magnetism and pinning effects. Furthermore, the metamagnetic transition understood to be due to the overwhelming α-$Fe_2O_3$ phase was observed in 5, 7, and 9 hours annealed samples while it showed significance only in the latter. This study underscores a simple and cost-effective means for controlling the structural and magnetic properties of iron oxide nanostructures by relative phase control.

**Acknowledgments**

The research was supported by the US Department of Energy, Office of Basic Energy Sciences, Division of Materials Science and Engineering under Award No. DE-FG02-07ER46438.